\documentclass{revtex4}

\usepackage{graphicx}
\begin{document}
% You should use BibTeX and revtex.bst for references
% \bibliographystyle{revtex}
\bibliographystyle{apsrev}

% Use the \preprint command to place your local institutional report
% number  and your conference paper identification number on the
% title page in preprint mode. Multiple \preprint commands are allowed.
%\preprint{}
% \preprint{SNOWMASS E3038}
% \preprint{SCIPP 01/34}

\begin{minipage}[t]{\textwidth}
\parbox[t]{0.1\textwidth}{
	\includegraphics[height=5cm,angle=0]{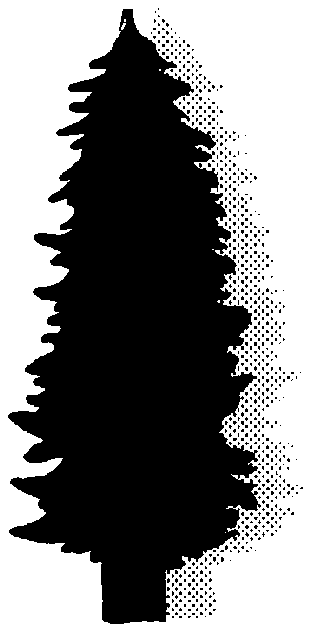}%
	}
\hfill
\raisebox{5cm}{
  \parbox[t]{0.5\textwidth}{\hspace*{\fill} SCIPP 01/34 \newline
                            \hspace*{\fill} SNOWMASS E3038}
}
\end{minipage}
\vfill

%Title of paper
\title{Vertex Detection for a Charm Tag in $e^+e^- \rightarrow W^+W^-$
	at a High Energy Electron-Positron Linear Collider}
% Optional argument for running titles on pages
%\title[]{}

% repeat the \author .. \affiliation  etc. as needed
% \email, \thanks, \homepage, \altaffiliation all apply to the current
% author. Explanatory text should go in the []'s, actual e-mail
% address or url should go in the {}'s for \email and \homepage.
% Please use the appropriate macro for the type of information

% \affiliation command applies to all authors since the last
% \affiliation command. The \affiliation command should follow the
% other information

\author{Wolfgang Walkowiak}
\email[]{walkowia@scipp.ucsc.edu}
%\homepage[]{Your web page}
%\thanks{}
%\altaffiliation{}
\affiliation{Santa Cruz Institute for Particle Physics \\
	University of California, Santa Cruz \\
	1156 High Street \\
	Santa Cruz, CA 94064}

%Collaboration name if desired (requires use of superscriptaddress
%option in \documentclass). \noaffiliation is required (may also be
%used with the \author command).
%\collaboration{}
%\noaffiliation

\date{\today}

\begin{abstract}
% insert abstract here
The study of the process $e^+e^- \rightarrow W^+W^-$ at Linear
Collider energies presents a good opportunity to investigate anomalous
triple gauge boson couplings and $W^+_LW^-_L$ rescattering.  The
helicity analysis of the $e^+e^- \rightarrow W^+_LW^-_L$ decays will
benefit if the charm quark containing jet can be identified for events
which contain one hadronic $W$ boson decay to a charm and another quark.  
A JAVA implementation of the SLD collaboration's topological vertex
finding algorithm (ZVTOP) in the linear collider analysis framework
has been used to extract charm tag efficiencies and purities based on
vertex multiplicities.
\end{abstract}

% insert suggested PACS numbers in braces on next line
% \pacs{}

%\maketitle must follow title, authors, abstract and \pacs
\maketitle
\vfill
\begin{center}
{\small Talk presented at the 2001 Snowmass Workshop on the
Future of Particle Physics \\}
{July 1 -- July 20, 2001}
\vfill
{\small This work was supported by the Department of Energy,
	Grant \#DE-FG03-92ER40689.}
\end{center}

\vfill
\newpage
\setcounter{page}{0}

% body of paper here - Use proper section commands
% References should be done using the \cite, \ref, and \label commands
% 
%\section{}
%\label{}
%\subsection{}
%\subsubsection{}

\section{Motivation}

The next $e^+e^-$ linear collider will provide a good opportunity to
study strong electroweak symmetry breaking in the process 
$e^+e^- \rightarrow W^+W^-$, which is predominant at center-of-mass
energies of 500~GeV and higher.  Strong electroweak symmetry breaking
is expected to be seen as deviations from the standard model in two
possible ways:  either anomalous couplings at the $W^+W^-\gamma$ and 
$W^+W^-Z$ vertices will be introduced or $W^+_LW^-_L$ final state 
rescattering effects may occur \cite{E3038_walkowiak_0714_barklow}.

Studies of strong electroweak symmetry breaking effects 
\cite{E3038_walkowiak_0714_barklow2} make use of a 
helicity angle analysis technique for the $W^+W^-$ state and employ a
maximum likelihood method to fit alternatively for the two
coefficients describing anomalous couplings or the complex form factor
$F_T$ in the case of final state rescattering. These analyses are
usually carried out with one $W$ decaying leptonically, the other
hadronically.  In the absence of flavor tagging, the latter decay
introduces an ambiguity in the measurement of 
two of the five helicity angles entering
the maximum likelihood fit.  The determination of the flavor of one of
the two hadronic jets will enhance the sensitivity by an equivalent
luminosity gain of up to a factor 1.8 as shown in
Figure~\ref{E3038_walkowiak_0714:fig1}.  The study presented here attempts
to quantify the possible gain if flavor tagging is
employed in $e^+e^-\rightarrow W^+W^-$ decays using the current linear
collider detector (LCD) design models as suggested in
\cite{E3038_walkowiak_0714_wwewsb}.

\section{Monte Carlo Fast Simulation and Analysis Technique}

Monte Carlo Data Samples for the process $e^+e^-\rightarrow W^+_LW^-_L$ 
of 10000 events each were generated with the generator PANDORA-PYTHIA 
\cite{E3038_walkowiak_0714_pandora,E3038_walkowiak_0714_ppy}
for three center-of-mass system (CMS) energies at 500~GeV, 1000~GeV
and 1500~GeV and for both LCD options
\cite{E3038_walkowiak_0714_brau1,E3038_walkowiak_0714_bruce1}, 
the large gaseous detector (LD) and the Silicon detector (SD).  
The events generated with a TechniRho of 1600~GeV mass according
to the model described in \cite{E3038_walkowiak_0714_iddir}
have a more central polar angle distribution
than Standard Model $e^+e^-\rightarrow W^+W^-$ events
because of their enhanced $F_T$ amplitude.  
Only events with one $W$ decaying leptonically and the other
hadronically are produced. 
Fast Monte Carlo track smearing is applied using the JAVA based LCD
analysis software \cite{E3038_walkowiak_0714_wwmcfast}.

In order to not confuse hadrons stemming from different $W$s, only
events with am electron or muon in the leptonic $W$ decay are
selected. The $W$ production angle is restricted to $|\cos
\Theta^*_W| < 0.90$ yielding typically 6000 to 6500 events. 
After the charged lepton track has been removed the DURHAM jetfinder 
\cite{E3038_walkowiak_0714_durham} is employed to divide the event
into two jets.  The smeared charged tracks and neutral particle
vectors created from Monte Carlo truth information are used as input
to the jetfinder.  
Each jet is associated with one of the primary $W$ decay quarks by
angular proximity of the respective momentum vectors.  According to
the quark type the jet is associated with, jets are classified as 
either up-type or down-type jets.

\begin{figure}[htb]
  \hfill
  \parbox[t]{0.45\textwidth}{
  \includegraphics[width=0.45\textwidth]{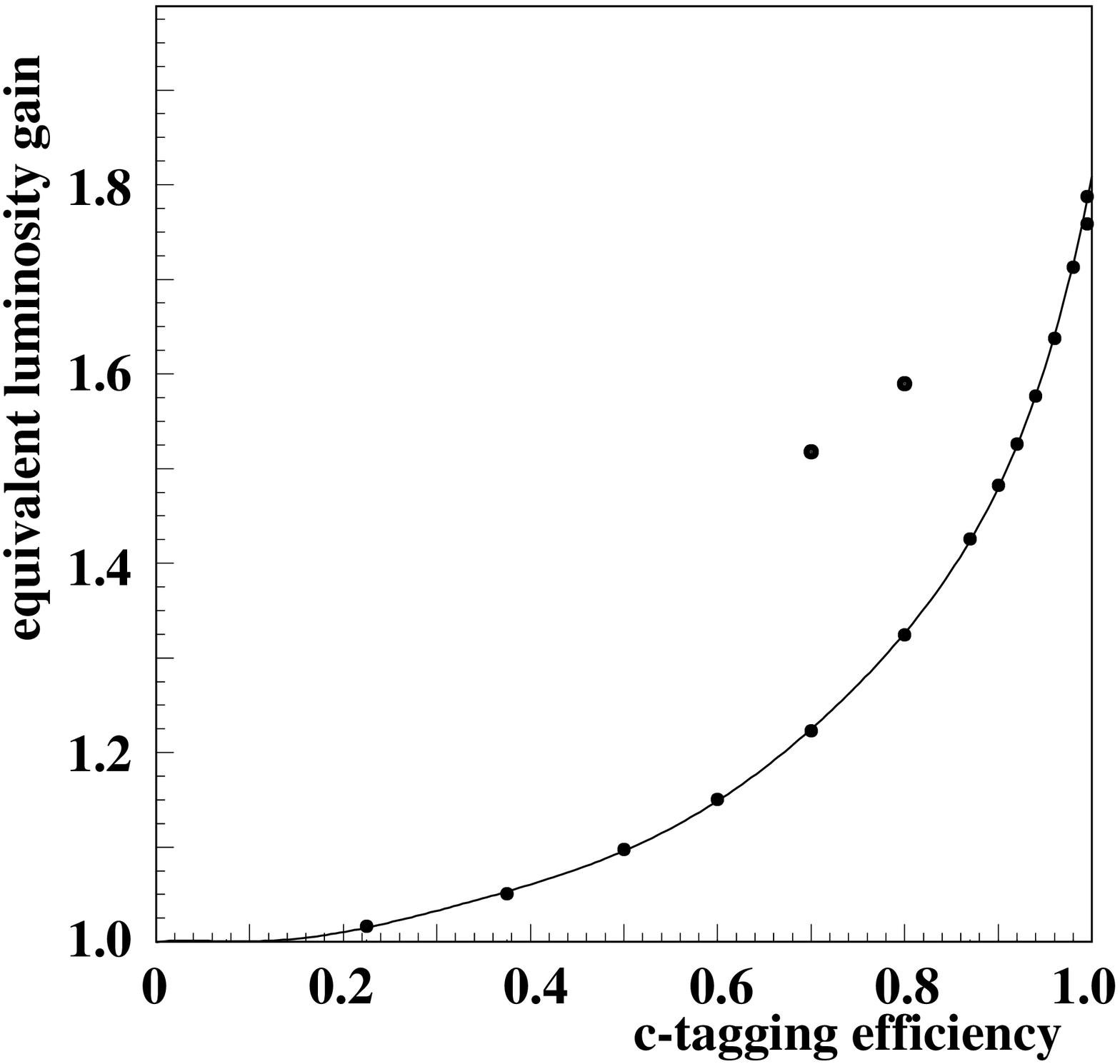}%
  \caption{Equivalent luminosity gain as a function of the charm
	tag efficiency \cite{E3038_walkowiak_0714_barklow2}.  The
	tagging efficiency includes possible mistags.  (The two 
	separate data points above the curve are obtained for the
	low energy theorem (LET) limit.)}
  \label{E3038_walkowiak_0714:fig1}
  }
  \hfill
  \parbox[t]{0.45\textwidth}{
  \includegraphics[width=0.45\textwidth]{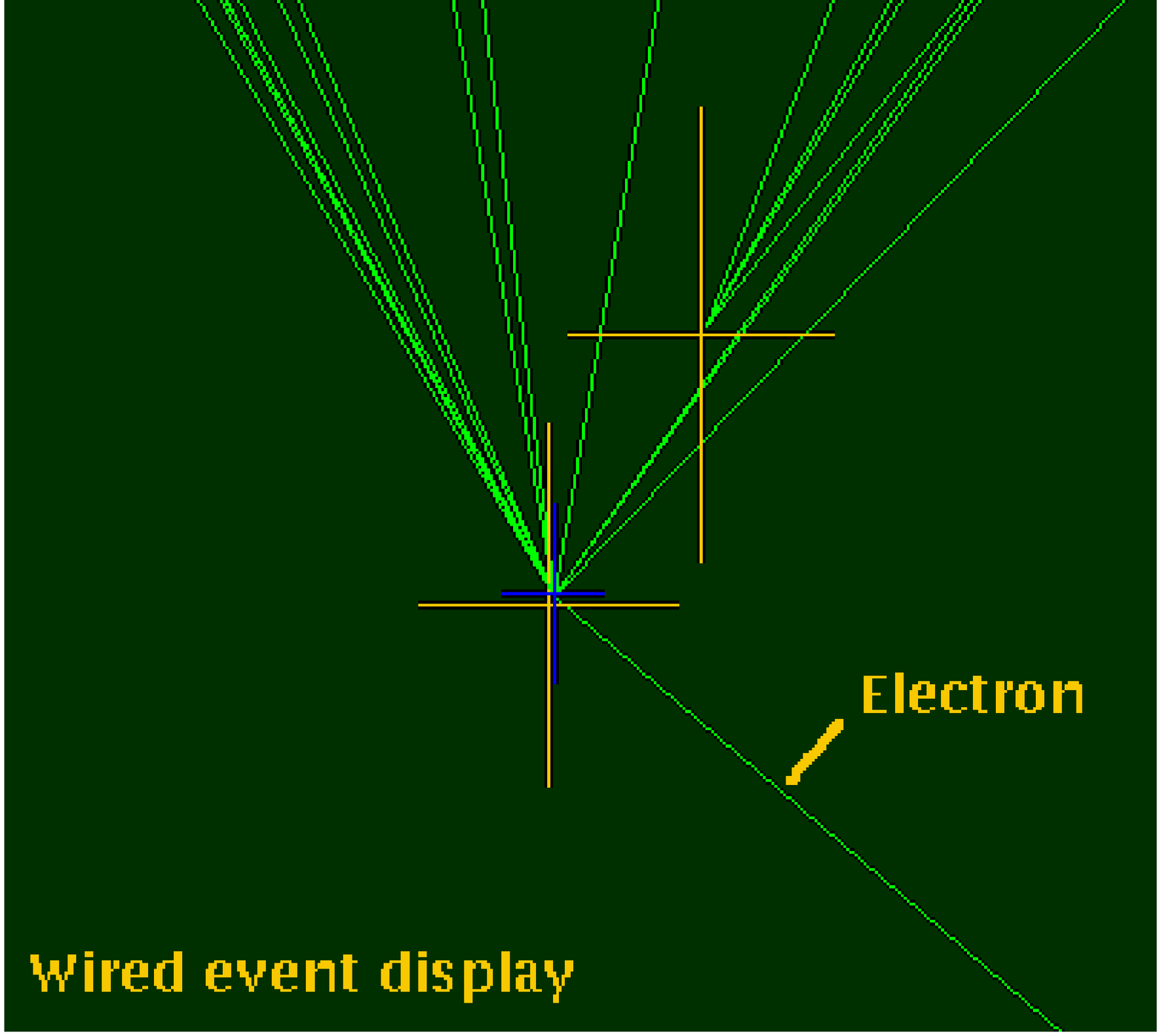}%
  %% \vspace{0.5cm}
  \caption{Displaced charm vertex as reconstructed by ZvTopVertexer
  (side view). 
	%% Typical sigmas for the reconstructed coordinates 
	%% are 50 to 100~$\mu$m.
	The coordinate errors shown are enlarged by a factor 100.}
  \label{E3038_walkowiak_0714:fig2}
  }
  \hfill
\end{figure}

Finally, ZvTopVertexer \cite{E3038_walkowiak_0714_wwmcfast}, the JAVA
implementation of SLD's topological vertexing algorithm
\cite{E3038_walkowiak_0714_jackson1}, is used to reconstruct vertices
from the charged tracks of each jet (see Figure 
\ref{E3038_walkowiak_0714:fig2}).  Since the multiplicity of vertices
found in a jet
depends on the lifetime of the $W$ decay particles, jets induced by a
charm quark are expected to have higher vertex multiplicities.  The
simple charm tag applied in this study requires at least two
reconstructed vertices $N_{vert}$ for the jet in question after
rejecting $K^0_S$ decay vertices by a cut on the invariant vertex
mass ($|m_{vtx}-m_{K^0_S}| < 25$~MeV) applied to the furthest outlying
vertex.

\begin{figure}[htb]
  \vspace*{-0.9cm}
  \hfill
  \parbox[t]{0.45\textwidth}{
  \includegraphics[width=0.40\textwidth]{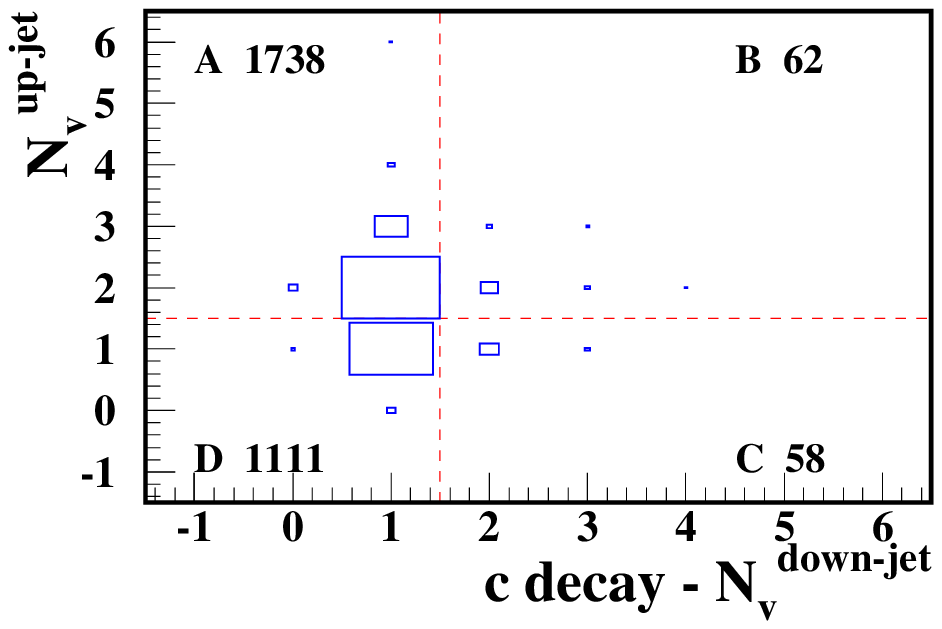}%
  \vspace{-0.5cm}
  \caption{Number of vertices of up-quark type jets vs. number of vertices
	of down-quark type jets for events with a $W\rightarrow cx$ decay. 
  	This example is for $E_{CMS} = 500$~GeV and the LD detector design.}
  \label{E3038_walkowiak_0714:fig3}
  }
  \hfill
  \parbox[t]{0.45\textwidth}{
  \includegraphics[width=0.40\textwidth]{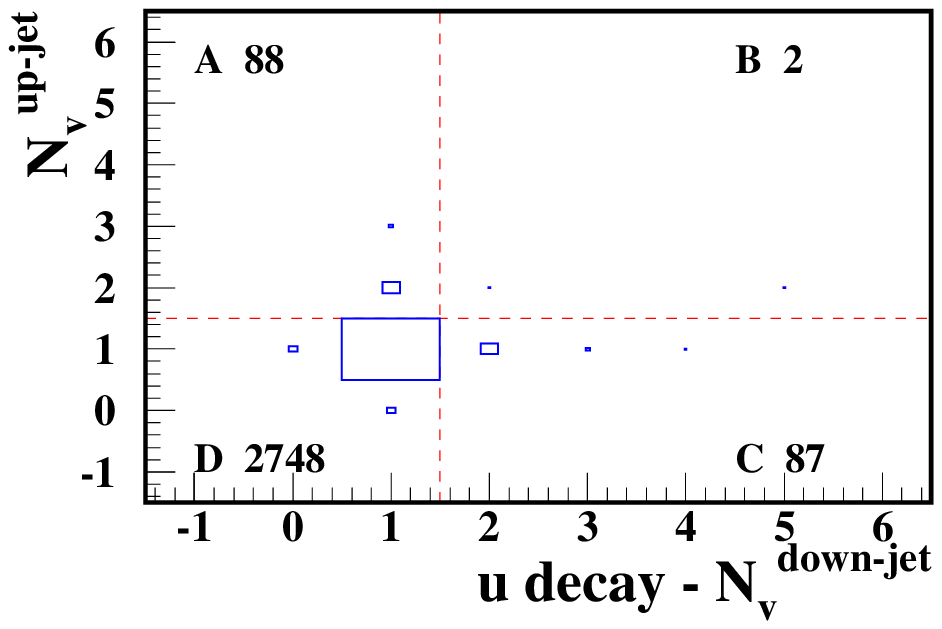}%
  \vspace{-0.5cm}
  \caption{Number of vertices of up-quark type jets vs. number of vertices
	of down-quark type jets for events without a $W\rightarrow cx$ decay. 
  	This example is for $E_{CMS} = 500$~GeV and the LD detector design.}
  \label{E3038_walkowiak_0714:fig4}
  }
  \hfill
\end{figure}

Figures \ref{E3038_walkowiak_0714:fig3} and
\ref{E3038_walkowiak_0714:fig4} show the distributions of
reconstructed vertices per jet and their correlation for up-type and
down-type jets for events with a $W\rightarrow cx$ decay and without
it, respectively. For the calculation of the charm tag efficiencies
and purities, Field A in Figure \ref{E3038_walkowiak_0714:fig3} 
contains correctly tagged events, while events in field C and in the 
fields A, B and C in Figure \ref{E3038_walkowiak_0714:fig4} are considered
mistags.  In the case of $W\rightarrow cx$ events in which both
jets have two or more vertices (field B in Figure 
\ref{E3038_walkowiak_0714:fig3}) we assume a 50\% chance to select the
correct jet.

\section{Results}

\begin{table}[htb]
  \hfill
  \parbox[t]{0.45\textwidth}{
  \begin{tabular}{|c|ccc|}
    \hline
    LC $E_{CMS}$ & 500~GeV & 1000~GeV & 1500~GeV \\
    \hline
    $\varepsilon_c$ &  $59.6\pm0.9$\% & $61.2\pm0.8$\% & $62.6\pm0.8$\% \\
    $p_c$           &  $86.9\pm0.4$\% & $89.6\pm0.4$\% & $92.1\pm0.3$\% \\ 
    $A$             &          73.9\% &         79.3\% &         84.3\% \\
    $Q$             &          32.5\% &         38.5\% &         44.4\% \\
    \hline
  \end{tabular}
  \caption{Results on the c-tag efficiency $\varepsilon_c$ and purity
    $p_c$, the analyzing power $A=2\,p_c-1$ and the effective tagging
    efficiency $Q=\varepsilon_c\,A^2$ for the LD detector design.}
  \label{E3038_walkowiak_0714:tab1}
  }
  \hfill
  \parbox[t]{0.45\textwidth}{
  \begin{tabular}{|c|ccc|}
    \hline
    LC $E_{CMS}$ & 500~GeV & 1000~GeV & 1500~GeV \\
    \hline
    $\varepsilon_c$ &  $60.1\pm0.9$\% & $61.5\pm0.8$\% & $62.5\pm0.8$\% \\
    $p_c$           &  $87.8\pm0.4$\% & $89.6\pm0.4$\% & $90.9\pm0.4$\% \\ 
    $A$             &          75.5\% &         79.2\% &         81.8\% \\
    $Q$             &          34.3\% &         38.6\% &         41.8\% \\
    \hline
  \end{tabular}
  \caption{Results on the c-tag efficiency $\varepsilon_c$ and purity
    $p_c$, the analyzing power $A=2\,p_c-1$ and the effective tagging
    efficiency $Q=\varepsilon_c\,A^2$ for the SD detector design.}
  \label{E3038_walkowiak_0714:tab2}
  }
  \hfill
\end{table}

\begin{figure}[htb]
  \vspace*{-0.9cm}
  \hfill
  \parbox[t]{0.45\textwidth}{
  \includegraphics[width=0.45\textwidth]{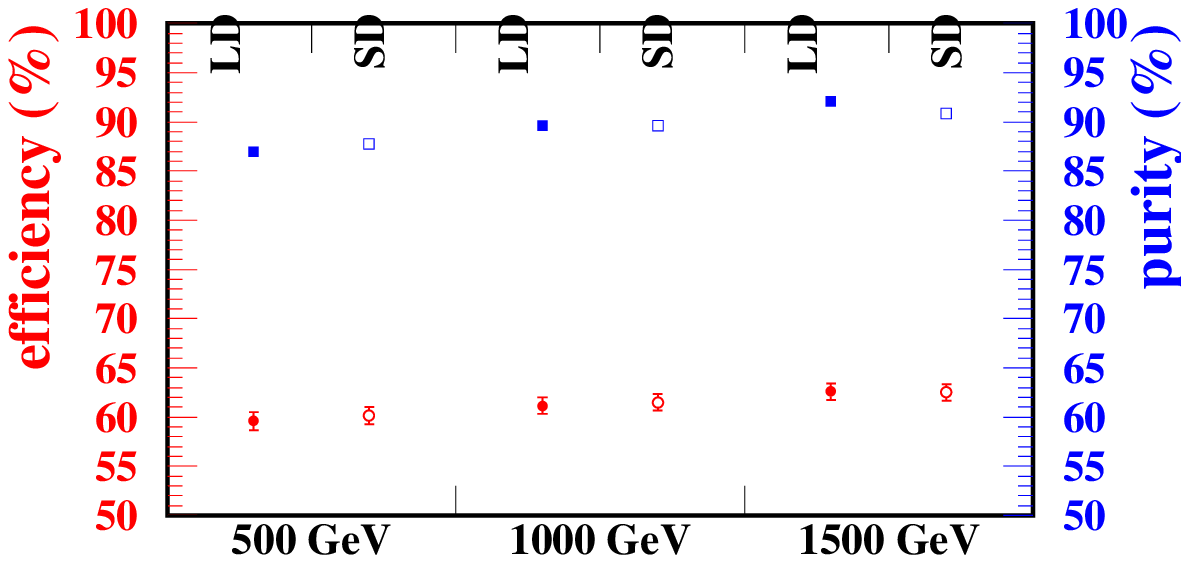}%
  \vspace{-1.0cm}
  \caption{C-tag efficiency (lower red circles) and purity (upper blue
  	squares) without the $K^0_S$ vertex veto for LD and SD designs 
	at various CMS energies.}
  \label{E3038_walkowiak_0714:fig5}
  }
  \hfill
  \parbox[t]{0.45\textwidth}{
  \includegraphics[width=0.45\textwidth]{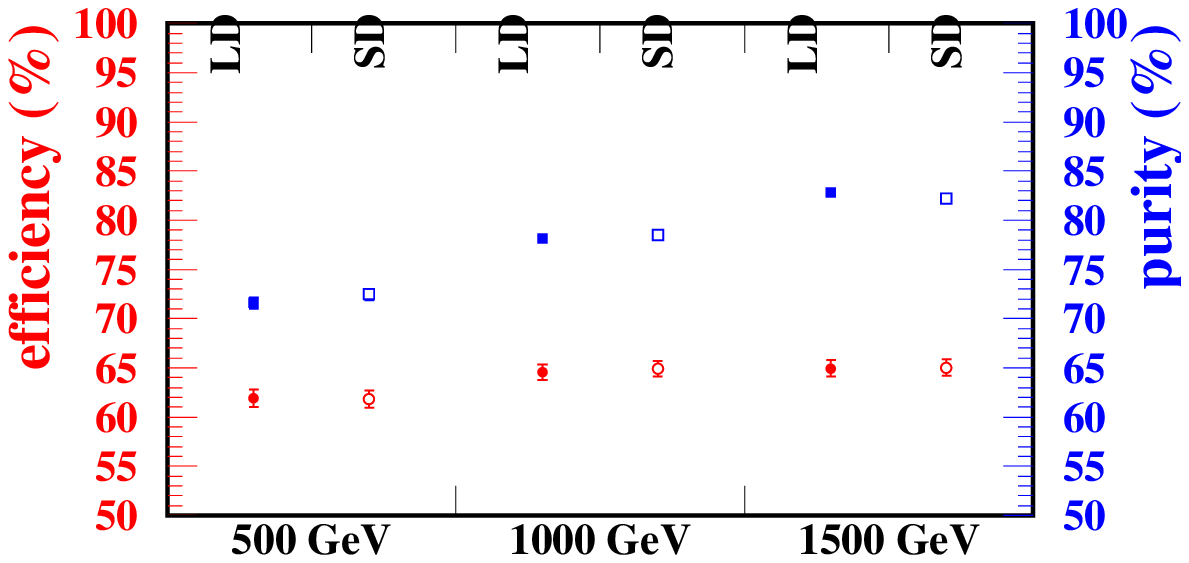}%
  \vspace{-1.0cm}
  \caption{C-tag efficiency (lower red circles) and purity (upper blue
  	squares) without the $K^0_S$ vertex veto for LD and SD designs 
	at various CMS energies.}
  \label{E3038_walkowiak_0714:fig6}
  }
  \hfill
\end{figure}

The resulting efficiencies and purities for correctly tagged jets for
event samples at three different LC CMS energies 
are collected in
Table~\ref{E3038_walkowiak_0714:tab1} and 
Table~\ref{E3038_walkowiak_0714:tab2} for the LD and SD design
options, respectively.  The efficiencies at a level 61\% and
the purities at about 90\% only show a slight dependence on the CMS energy.  
The latter causes the increase of the effective tagging power $Q$ with energy
from about 33\% to 43\%.
As can be seen in Figure \ref{E3038_walkowiak_0714:fig5} only minor
differences between the two LCD design options exist.  
Figure \ref{E3038_walkowiak_0714:fig6} points to the importance of the
$K^0_S$ decay vertex veto in order to reach a high purity level.  While the
charm tag efficiency is only decreased by a few percent, the purity
is enhanced by 10\% or more.  

Achieving charm tag efficiencies of approximately 61\% and purities of 
approximately 90\%
(in absence of other backgrounds than $W\rightarrow ux$ decays) with a
simple vertex multiplicity tag is encouraging.  However, in order to
obtain the effective charm tagging efficiency nescessary to substantially
increase the sensitivity of the helicity measurement, additional
measures to enhance the charm  vertex reconstruction as well as
the use of additional discriminating variables, possibly in a neural
net based technique, will be required.

% \end{table}
% figures should be put into the text as floats.
% Use the graphicx package (distributed with LaTeX2e).
% See the LaTeX Graphics Companion by Michel Goosens, Sebastian Rahtz,
% and Frank Mittelbach for instance.
%
% Here is an example of the general form of a figure:
% Fill in the caption in the braces of the \caption{} command. Put the label
% that you will use with \ref{} command in the braces of the \label{} command.
%
% \begin{figure}
% \includegraphics{}%
% \caption{}
% \label{}
% \end{figure}

% tables follow here or maybe be put in the text
%
% Here is an example of the general form of a table:
% Fill in the caption in the braces of the \caption{} command. Put the label
% that you will use with \ref{} command in the braces of the \label{} command.
% Insert the column specifiers (l, r, c, d, etc.) in the empty braces of the
% \begin{tabular}{} command.
%
% \begin{table}
% \caption{}
% \label{}
% \begin{tabular}{}
% \end{tabular}
% \end{table}

% If you have acknowledgments, this puts in the proper section head.

\begin{acknowledgments}
I would like to thank Tim Barklow, Gary Bower, Michael Peskin,
Bruce Schumm and Abi Soffer for stimulating discussions and
suggestions as well as the organizers and
participants of Snowmass 2001 for a successful workshop. 
\end{acknowledgments}

% Create the reference section using BibTeX:
\bibliography{E3038_walkowiak_0714.bib}

\end{document}